\documentclass[12pt, draftclsnofoot, onecolumn]{IEEEtran}

\usepackage{cite}
\usepackage[hyphens]{url}
\usepackage[hidelinks]{hyperref}
\hypersetup{breaklinks=true}
\usepackage{amsmath}
\usepackage{amssymb}
\usepackage[dvips]{graphicx}
\usepackage{setspace}
\usepackage{epsfig}
\usepackage{subfigure}
\usepackage{color}

\newcommand{\figsize}{0.37}
\newcommand{\ssp}{\text{sp}}
\newcommand{\diag}{\text{diag}}

\newtheorem{theo}{Theorem}

\newtheorem{rem}{Remark}
\newtheorem{prop}{Proposition}
\newtheorem{definition}{Definition}

\begin{document}

\title{On the Energy and Data Storage Management in Energy Harvesting Wireless Communications}

\author{\IEEEauthorblockN{Sami Ak{\i}n and M. Cenk Gursoy}\\
\thanks{S. Ak{\i}n is with the Institute of Communications Technology, Leibniz Universit\"at Hannover, 30167 Hanover, Germany, (E-mail: sami.akin@ikt.uni-hannover.de).}
\thanks{M. C. Gursoy is with the Department of Electrical Engineering and Computer Science, Syracuse University, Syracuse, NY 13244 USA (E-mail: mcgursoy@syr.edu)}
\thanks{This work was supported by the German Research Foundation (DFG) -- FeelMaTyc (FI 1236/6-1)}}

\maketitle

\begin{abstract}
Energy harvesting (EH) in wireless communications has become the focus of recent transmission technology studies. Herein, energy storage modeling is one of the crucial design benchmarks that must be treated carefully. Understanding the energy storage dynamics and the throughput levels is essential especially for communication systems in which the performance depends solely on harvested energy. While energy outages should be avoided, energy overflows should also be prevented in order to utilize all harvested energy. Hence, a simple, yet comprehensive, analytical model that can represent the characteristics of a general class of EH wireless communication systems needs to be established. In this paper, invoking tools from large deviation theory along with Markov processes, a firm connection between the energy state of the battery and the data transmission process over a wireless channel is established for an EH transmitter. In particular, a simple exponential approximation for the energy overflow probability is formulated, with which the energy decay rate in the battery as a measure of energy usage is characterized. Then, projecting the energy outages and supplies on a Markov process, a discrete state model is established and an expression for the energy outage probability for given energy arrival and demand processes is provided. Finally, under energy overflow and outage constraints, the \emph{average data service (transmission) rate} over the wireless channel is obtained and the \emph{effective capacity} of the system, which characterizes the maximum data arrival rate at the transmitter buffer under quality-of-service (QoS) constraints imposed on the data buffer overflow probability, is derived.
\end{abstract}

\begin{IEEEkeywords}
Energy harvesting, energy overflow, energy outage, effective capacity, queueing theory, large deviation theory, Markov process.
\end{IEEEkeywords}

\section{Introduction}\label{introduction}
Due to the unprecedented growth in the number of wireless devices and systems, requiring an ever-increasing amount of energy, research efforts in recent years have focused on more sophisticated energy management techniques. In particular, energy harvesting (EH) technology has attracted significant interest from the research community as a means both to reduce the carbon footprint of communication networks and to provide increased autonomy to wireless devices \cite{energy_harvesting_ulukus_2015,ozel2015fundamental,gunduz2014designing,tadayon2013power}. The challenges in these studies arise from the complexity in modeling practical energy storage technologies, variable energy consumption patterns, and the stochastic aspects of natural energy sources, such as wind and solar. Researchers have employed various mathematical models to study energy storage mechanisms for EH systems. Among these models, queueing-theoretic energy quantization models \cite{tadayon2013power} are arguably the most common and well-established ones. These models can reflect the practical characteristics of energy storage systems relatively accurately, and are generally amenable to mathematical analysis when invoked in communication systems. Concurrently, understanding the energy consumption profile of wireless devices is also essential, as the energy retrieval rate from a storage unit may affect the lifetime of a device. However, while it was relatively easy to model and estimate the energy consumption behavior of a communication device in the past because mostly the circuit, baseband, radio frequency and power amplifier components consumed the energy, nowadays it is much more difficult to accurately understand and model the energy needs of communication devices due to the ever-increasing complexity of modern devices and applications \cite{rice2010measuring}. Therefore, along with the introduction of EH wireless communication technology, which relies on environment-friendly techniques to generate energy from renewable resources, the effective use of the generated energy to guarantee energy availability when required, led to a paradigm shift in research on radio resource allocation \cite{ahmed2015survey}. Particularly, in addition to spectral efficiency and quality-of-service (QoS) constraints, economic use of energy has emerged as another requirement. The concern lies in estimating the periodicity and magnitude of the exploited energy source, deciding which parameters to tune, and simultaneously avoiding premature energy depletion before the next recharge cycle \cite{sudevalayam2011energy}. Hence, the goal is to characterize a tradeoff between the performance levels and lifespan of built-in energy units. This new perspective compels the need to understand energy storage technologies and the associated performance levels.

\subsection{Related Work}
Since the natural energy sources are stochastic, the power and data management policies in EH wireless communication systems are different from their counterparts that depend on the grid energy or non-rechargeable batteries. Moreover, considering a system that has a data buffer as well as a battery, the optimal control of an EH wireless communication system requires managing the transmission rate by monitoring both the traffic load and the stored energy, and respecting the causality constraints in both the data and energy arrivals. The authors in \cite{ozel2011transmission} and \cite{6202352} controlled the transmit power levels subject to energy storage capacity and causality constraints, and introduced a directional offline water-filling algorithm that optimizes the delay-constrained throughput. Separately, the authors in \cite{ozel2011transmission} considered a transmitter model with a battery and a data buffer. Assuming that both data and harvested energy packets randomly arrive at the data buffer and the battery, respectively, the authors in \cite{yang2012optimal} developed optimal offline scheduling policies to adaptively change transmission rates under a deterministic system setting. The authors in \cite{6144764} and \cite{6253061} extended this analysis to a transmitter with finite-capacity battery, where the latter also considered energy leakage at the transmitter. The authors in \cite{5992840} considered the offline minimization of transmission completion time in broadcast links under EH constraints.

A hybrid energy storage system model with one unlimited battery and a limited super-capacitor is considered in \cite{6620534}, where the authors maximize the throughput under a data transmission deadline constraint. Moreover, having a transmitter utilizing both the harvested energy and the grid energy, the authors in \cite{liu2015delay} provided an analysis on the average data queueing delay and the average power consumption from the grid by formulating the data queueing and energy storage as a two-dimensional Markov chain. The authors in \cite{arafa2015optimal} and the authors in \cite{arafa2017energy} investigated the effects of decoding and processing costs in one-way and two-way channels, respectively, where the transmitter needs to adjust its transmission power policies regarding not only its own energy values but also the receiver's energy values.

One more fundamental consideration in using certain EH sources, which is different from those in using non-rechargeable batteries, is the maximum rate at which we can utilize the harvested energy \cite{Kansal:2007:PME:1274858.1274870}. Therefore, we have to capture the uncertainty in not only the energy source but also the consumption. With this motivation in mind, the authors in \cite{993375} modeled a battery similarly to a server with a finite service capacity, and the data packets similarly to customers to be served, and analyzed the performance using a queueing-theoretic approach. This work is one of the earlier attempts to model the battery using queueing theory. Regarding a sensor node as a paired queueing system with two buffers, one for the accumulated energy and the other for the arriving data, Cuypere \emph{et al.} analyzed its performance numerically, and investigated the energy-information tradeoff \cite{3229100}. A more comprehensive and practical, yet simple enough, model, which also takes the variations in harvested energy into consideration, appeared in \cite{Kansal:2007:PME:1274858.1274870} and \cite{Kansal:2004:PAT:1012888.1005714}. We can easily incorporate the aforementioned queueing model to a general class of stochastic energy sources. Bounding the energy arrival from an energy source with lower and upper bounds, and bounding the energy consumption with an upper bound, they showed that a device can operate forever as long as it has a storage capacity greater than the total burstiness defined by the upper and lower bounds \cite{Kansal:2004:PAT:1012888.1005714}. They provided the stability condition such that the average energy arrival rate is less than or equal to the average energy demanded for consumption. Although these deterministic bounds provide a framework to study the steady-state behavior of energy production and consumption processes, they indeed capture the worst-case scenario. Also, these bounds do not take advantage of the statistical nature of the energy arrival and consumption processes. Separately, considering statistical bounds, Srivastava \emph{et al.} invoked large deviation theory and showed that the energy underflow probability, i.e., the probability that the energy level in the battery is below a defined threshold, scales exponentially as a function of the battery size and a constant in the asymptotic regime of large battery size \cite{srivastava2013basic}. Finally, the authors in \cite{jia2017data} characterized the average backlog for both constant and random data arrivals at a finite-size data buffer considering a Bernoulli energy arrival process.

Last but not least, considering a network of energy harvesters, namely energy packet networks with energy harvesting, the authors in \cite{gelenbe2016energy} and the ones in \cite{kadioglu2018product} invoked a branch of queuing theory called G-networks and queueing networks with product-form solution, respectively, in order to compute relevant performance metrics of such networks operating with intermittent energy. Another research direction in energy harvesting wireless communication headed the simultaneous wireless information and power transfer. We refer interested readers to \cite{tran2018resource, 8629017, hu2018swipt, 8620264}. Finally, we refer to studies where energy harvesting is investigated in multiple-input multiple-output systems \cite{7790901, le2018joint, 8629017, Le2019}.

\subsection{Contributions}
In EH communications, storage modeling is one of the crucial design benchmarks that we have to treat diligently. Especially in communication systems whose operations depend solely on the harvested energy, understanding the storage dynamics becomes fundamental from a design viewpoint. While a tenable energy storage is of paramount importance for us to avoid energy outages, preventing energy overflows due to the limited battery size is also necessary in order to fully utilize the harvested energy. Accordingly, we need a simple, yet comprehensive analytical model that can encapsulate the characteristics of a generic EH wireless communication system. In this paper, we introduce an analytical framework that system designers can use in order to understand the performance levels in a general class of EH communication systems under energy overflow and outage probability constraints, and QoS requirements. Similarly to \cite{srivastava2013basic} and \cite{zhang2015joint}, we take advantage of large deviation theory and Markov processes; however, different from these studies, we consider simultaneously the energy overflow and outage probabilities in an EH communication system, and perform the throughput analysis. Specifically, while \cite{srivastava2013basic} provides the energy underflow probability in the battery using large deviation theory , the authors in \cite{zhang2015joint} maximize the effective capacity under the energy underflow probability constraint. On the other hand, we formulate the energy overflow probability in the battery as an exponential function of the battery size by invoking large deviation theory, and the energy outage probability by employing the Markov process analysis. Particularly, we consider that the transmitter initially sets a transmission power policy taking into account the energy overflow and outage probability constraints, and the energy arrival statistics. It sets the transmission rate based on the power allocation policy. We assume that the transmitter knows the channel statistics, but is unaware of the instantaneous channel state. Specifically, we have the following contributions:
\begin{enumerate}
	\item Using large deviation theory and queueing theory, we formulate a simple exponential approximation for the energy overflow probability, where we characterize the energy decay rate (i.e., the decay rate of the tail distribution of the stored energy) in the battery as a measure of energy utilization.
	\item Mapping the system evolution to a Markov chain, in which we have one state representing the energy outage, and other states representing the number of time frames since the last energy outage event, we provide an expression for the energy outage probability.
	\item Under energy overflow and outage constraints, we obtain the average data service rate in the wireless channel. Subsequently, we identify the effective capacity of the system, which characterizes the maximum data arrival rate at the transmitter buffer when there are QoS requirements in the form of constraints on the buffer overflow probability.
\end{enumerate}

Apart from this paper, we invoke large deviation theory and queueing theory in \cite{akin2017energy,akin2018energy}, where we characterize the energy underflow probability. Specifically in this paper, we have a focus primarily on controlling energy waste and then energy outages under QoS constraints, whereas in \cite{akin2017energy,akin2018energy}, we principally control energy underflows, and hence outages, in communication settings under stricter QoS constraints, where transmission interruptions are not tolerated at any time.

The rest of the paper is organized as follows. We introduce the transmission system model consisting of an EH device, a storage unit, and a data buffer in Section \ref{sec:system_model}. We characterize the energy storage performance measures such as energy overflow and energy outage in Section \ref{sec:Energy_Storage_Characterization}. We provide the throughput analysis in Section \ref{sec:transmission_throughput}, i.e., formulate the average data service rate in the channel in Subsection \ref{sec:reliable_throughput} and the effective capacity in Subsection \ref{sec:effective_capacity}. We present the numerical results in Section \ref{sec:numerical_results}. Our conclusions are provided in Section \ref{sec:conclusion}. We relegate the proofs to the Appendix.

\section{System Model}\label{sec:system_model}
\begin{figure}
	\centering
	\includegraphics[width=0.55\textwidth]{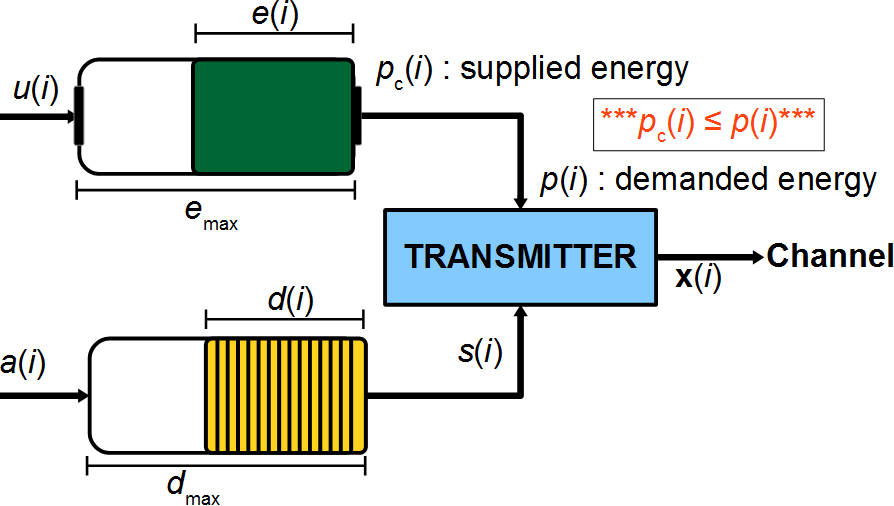}
	\caption{EH transmitter model consisting of a data buffer and a battery acting as an energy buffer.}\label{res:fig_2}
	\vspace{-0.0cm}
\end{figure}

We consider a discrete-time system model consisting of two separate queues at the transmitter corresponding to the energy and data buffers, respectively. Please, see Fig. \ref{res:fig_2} for an illustration of the system model. We describe each of these components separately.

\subsection{Energy Harvesting and Storage}\label{sec:Energy_Harvesting_Storage}
We denote the amount of the harvested energy in the $i^{\text{th}}$ time frame by $u(i)$ and the amount of energy demanded for data transmission by $p(i)$ for $i=1,2,\cdots$. We assume that $u(i)$ is stochastic and varies from one time frame to another, whereas the amount of energy demanded, $p(i)$, is a stochastic process\footnote{One can consider adaptive power control and modulation schemes, where transmitters adjust their power and the data modulation according to the signal-to-noise ratio at the corresponding receivers. Since the received signal-to-noise ratio at a receiver is a function of channel fading, the adjusted power levels, hence the energy demands, become stochastic.}, and its parameters\footnote{With parameters, we refer to distribution, mean, variance, and maximum and minimum values, etc.} are pre-determined by the transmitter. When setting $p(i)$, the transmitter regards certain constraints and performance requirements, e.g., energy overflow and outage probability constraints, which will be defined below. The energy level in the battery at the end of the $i^{\text{th}}$ time frame, denoted by $e(i)$, $i\in\mathbb{Z}^{+}$, is governed by the following update rule:
\begin{equation}\label{actual_capacity_2}
e(i)=\min\{[e(i-1)+u(i)-p(i)]^{+},e_{\max}\},
\end{equation}
where $[\cdot]^{+}\triangleq\max\{0,\cdot\}$, $e_{\max}$ denotes the battery capacity, and $e(0)$ is the initial energy level of the battery. As seen from (\ref{actual_capacity_2}), the energy harvested in the $i^{\text{th}}$ time frame, $u(i)$, is ready for consumption in the same time frame. If $e(i-1)+u(i)-p(i)>e_{\max}$, we will have an \emph{energy overflow} in the $i^{\text{th}}$ time frame, wasting some of the harvested energy. On the other hand, when $e(i-1)+u(i)<p(i)$, we will have an \emph{energy outage}.

While $p(i)$ is the rate of the energy demanded by the transmitter, the battery may not be always able to provide energy at this level. When there is enough energy in the battery, i.e., $e(i-1)+u(i)\geq p(i)$, the battery satisfies the energy demand completely. Otherwise, the battery provides what is left in it for the transmission of data. We denote the actual consumed energy in the $i^{\text{th}}$ time frame by $p_{\text{c}}(i)$, where
\begin{equation}\label{eq:power_level}
	p_{\text{c}}(i) = \begin{cases}
		p(i), &\text{if  $e(i-1)+u(i)\geq p(i)$},\\
		e(i-1)+u(i), &\text{otherwise}.
	\end{cases}
\end{equation}
Moreover, because the transmitter cannot spend more energy than what has been generated (i.e., energy causality), we have the following causality constraints:
\begin{equation}\label{energy_causality}
e(0)+U(t)\geq P_{c}(t),\quad\forall t,
\end{equation}
where $U(t)\triangleq\sum_{i=1}^{t}u(i)$ and $P_{c}(t)\triangleq\sum_{i=1}^{t}p_{\text{c}}(i)$ are the total energy harvested and consumed in the first $t$ time frames, respectively. Here, we assume that the transmitter knows the battery state perfectly.

\subsection{Data Buffer}
In the $i^{\text{th}}$ frame, $a(i)$ bits of data arrive at the transmitter, and is stored in the transmitter buffer. The buffer has a capacity of $d_{\max}$ bits, and the number of bits in the buffer in the $i^{\text{th}}$ time frame is $d(i)$. When a packet of data is transmitted and decoded by the receiver correctly, the data packet is removed from the transmitter buffer, and it is considered to be served. Let $s(i)$ denote the amount of data served in the $i^{\text{th}}$ frame. Assuming that the transmitter sends $r(i)$ bits over the channel in the $i^{\text{th}}$ time frame, we can clearly see that $s(i)=r(i)$ if the receiver is able to decode the transmitted data, and $s(i)=0$ otherwise. In order to ensure the delivery of the data, we assume that a simple automatic-repeat-request (ARQ) mechanism exists to acknowledge a successful reception, or to trigger the retransmission of erroneously decoded data. We assume that the transmitter does not have the channel state information, but knows the channel statistics. If the number of transmitted bits is smaller than the maximum rate that can be supported by the channel capacity in the $i^{\text{th}}$ time frame, transmission is assumed to be successful; otherwise, a decoding error occurs.

\section{Energy Storage Characterization}\label{sec:Energy_Storage_Characterization}
We explore the battery dynamics taking into consideration energy overflows and outages, and establish limits on the energy overflow and outage probabilities as the statistical battery (or energy management) constraints. Therefore, observing the structural similarity between a single server queuing system and the aforementioned storage model, we invoke queueing theory, large deviations theory \cite{chang_book} and network calculus \cite{jiang_book} to understand the energy overflows in the battery. Then, projecting subsequent energy demands that have been satisfied, and energy outages on a Markov process, we characterize the energy outage probability.

\subsection{Energy Overflow}\label{sec:energy_overflow}
We know that given a single service provider, the steady-state queue length tail distribution in a queueing system with a first in-first out policy, assuming it exists, has a characteristic decay rate, i.e., the decay rate of the tail probability of the queue \cite{chang1994effective}. When the queueing capacity is infinite, we obtain a simple exponential expression for the queue overflow probability (i.e., the probability that the queue is greater than a threshold), which is a function of the desired threshold and the characteristic decay rate. In practical systems, when the queueing capacity is large, this exponential expression approximates the queue overflow probability very closely. Likewise, for the EH transmitter model introduced in Section \ref{sec:system_model}, we assume that the battery size is infinite, i.e., $e_{\max}=\infty$, and define the decay rate of the tail distribution of the energy in the battery as follows.
\begin{definition}\label{def:definition_decay_rate}
Given a stationary and ergodic energy arrival process, $u(i)$, and a stationary and ergodic energy demand process, $p(i)$, under the stability condition\footnote{Note that while the energy causality constraint remains between the amount of the accumulated energy and the amount of the consumed energy, it is assumed that the stability condition, in which the average demanded energy is greater than the average accumulated energy, exists so that the amount of energy in the battery in the steady-state does not go to infinity; thus, we control the energy waste. On the other hand, if our primary concern is to control transmission interruptions, we can impose stability conditions to avoid battery being depleted. In particular, we set $\mathbb{E}\{p(i)\}<\mathbb{E}\{u(i)\}$ as the stability condition, similarly to \cite{srivastava2013basic,akin2017energy,akin2018energy}.}, i.e., $\mathbb{E}_{u}\left[u(i)\right]<\mathbb{E}_{p}\left[p(i)\right]$, where $\mathbb{E}\left[\cdot\right]$ is the expected value operator, the \emph{energy decay rate} of the battery is defined as
\begin{equation}\label{decay_rate}
\mu\triangleq-\lim_{e_{\text{th}}\to\infty}\frac{\ln\Pr\{e\geq e_{\text{th}}\}}{e_{\text{th}}},
\end{equation}
where $e_{\text{th}}$ is the desired energy level, and random variable $e$ corresponds to the steady-state distribution of the energy level in the battery.
\end{definition}

The definition of $\mu$ in (\ref{decay_rate}) suggests an approximation for the energy overflow probability in the steady-state given a large battery size\footnote{In certain practical scenarios, the battery capacity can be regarded as infinite with respect to the energy arrival and demand processes. For example, some transceivers require output power levels on the order of $2-100$ mW to communicate within a range of $30$ meters, while some solar cells produce $15$ mW/$\text{cm}^2$ \cite{safak2014wireless}, in which case an AAA alkaline battery \cite{website2} can be considered as having a very large capacity that can closely approximate infinite capacity.}, i.e., $\Pr\{e\geq e_{\text{th}}\}\approx \exp\big(-\mu e_{\text{th}}\big)$. In particular, the energy decay rate characterizes the exponential decay rate of the tail distribution of the energy level in the battery, and the exponential approximation holds when we have stationary and ergodic energy arrival and demand processes \cite{chang1995effective_conf}. With a large battery size and a target overflow probability, we can use the energy decay rate, $\mu$, as a tool to identify the probability that the energy level in the battery is above a defined threshold. The energy decay rate primarily depends on the energy arrival and demand processes. Large $\mu$ refers to an energy demand process that consumes the stored energy rapidly, while smaller $\mu$ means a moderate energy demand process.

For an infinite-size battery the instantaneous energy level is given by $e(i)=[e(i-1)+u(i)-p(i)]^{+}$. Moreover, because energy in our model is used for data transmission only, we further consider a work-conserving energy demand process\footnote{We consider that the transmitter always has data to transmit. Therefore, it consumes a certain amount of energy as long as there is energy in the battery, and we regard the energy demand process as work-conserving. We have this assumption because the harvested energy is utilized for data transmission only, and the control of energy overflows and outages becomes important for an efficient use of the harvested energy when there is data in the transmitter buffer. Otherwise, when there is no data in the buffer, the transmitter harvests energy until the battery becomes full, and then stops harvesting energy.} in the following analysis. Hence, noting that $u(i)$ and $p(i)$ are independent of each other, we have a unique $\mu^{\star}$ that satisfies \cite[Remark 9.1.2]{chang_book}
\begin{align}\label{eq:arrival_demand_balance}
\Lambda_{u}(\mu^{\star})+\Lambda_{p}(-\mu^{\star})=0,
\end{align}
where $\Lambda_{u}(\mu)\triangleq\lim_{t\to\infty}\frac{1}{t}\ln\mathbb{E}_{u}\left[\exp\left(\mu U(t)\right)\right]$ and $\Lambda_{p}(\mu)\triangleq\lim_{t\to\infty}\frac{1}{t}\ln\mathbb{E}_{p}\left[\exp\left(\mu P(t)\right)\right]$ are the G\"artner-Ellis limits, and are differentiable for $\mu\in\mathbb{R}$, when the moment generating functions $\mathbb{E}_{u}\left[\exp\left(\mu U(t)\right)\right]$ and $\mathbb{E}_{p}\left[\exp\left(\mu P(t)\right)\right]$ exist for $t>0$, respectively. $U(t)$ is the cumulative harvested energy as defined before, and $P(t)\triangleq\sum_{i=1}^{t}p(i)$. Noting that (\ref{eq:arrival_demand_balance}) holds for any stationary and ergodic energy arrival and demand processes with finite mean and variance, we provide the following two cases as examples in order to gain more insights.

\subsubsection{Constant energy demand}
Given an energy arrival process, let us assume that the energy demanded by the transmitter is constant over time. Then, we can easily find the following relation for a given energy decay rate, $\mu$: $p^{\star}=\lim_{t\to\infty}\frac{1}{t\mu}\ln\mathbb{E}_{u}\left[\exp\left(\mu U(t)\right)\right]$, where $p^{\star}$ is the minimum constant energy demand such that the steady-state energy overflow probability is $\Pr\{e(i)\geq e_{\text{th}}\}\approx\exp\big(-\mu e_{\text{th}}\big)$ for a given threshold value, $e_{\text{th}}$. In other words, targeting an energy decay rate, $\mu$, we should have the constant energy demand greater than or equal to $p^{\star}$ such that we can keep the steady-state energy overflow probability less than or equal to $\exp\big(-\mu e_{\text{th}}\big)$ for given $e_{\text{th}}$. We note that keeping the energy demand above $p^{\star}$ decreases the energy overflow probability while it causes an increase in the energy outage probability.

\begin{prop}\label{pro:mu_zero_and_infinity}
When the energy decay rate, $\mu$, goes to zero, the minimum constant energy demand goes to the average energy arrival level, $\mathbb{E}_{u}\left[u(i)\right]$. In particular,\[p^{\star}=\lim_{\mu\to0}\lim_{t\to\infty}\frac{1}{t\mu}\ln\mathbb{E}_{u}\left[\exp\left(\mu U(t)\right)\right]=\mathbb{E}_{u}\left[u(i)\right].\]Furthermore, when the energy decay rate, $\mu$, goes to infinity, the minimum constant energy demand goes to the maximum energy arrival level, $\max\{u(i)\}$. In particular, \[p^{\star}=\lim_{\mu\to\infty}\lim_{t\to\infty}\frac{1}{t\mu}\ln\mathbb{E}_{u}\left[\exp\left(\mu U(t)\right)\right]=\max\{u(i)\}.\]
\end{prop}
\emph{Proof}: See Appendix \ref{app:proposition_mu_zero_and_infinity}.

\subsubsection{Constant energy arrival}
Given an energy demand process, let us now assume that the energy arrival at the battery is constant\footnote{Although energy arrivals in general vary drastically in nature, recalling that our analytical framework works when both energy arrivals and demands are stochastic, ergodic and stationary processes with finite mean and variance given that the two processes are independent of each other, we consider a constant energy demand process as an example for mathematical tractability of (\ref{eq:arrival_demand_balance}). We refer interested readers to \cite{rajesh2011capacity, siddiqui2017performance} as well, where a practical sensor node prototype that assumes solar energy to be constant for optimal event detection probability is considered.} over time. Then, we can easily find the following relation for a given energy decay rate, $\mu$: $u^{\star}=-\lim_{t\to\infty}\frac{1}{t\mu}\ln\mathbb{E}_{p}\left[\exp\left(-\mu P(t)\right)\right]$, where $u^{\star}$ is the maximum constant energy arrival such that the steady-state energy overflow probability is $\Pr\{e\geq e_{\text{th}}\}\approx\exp\left(-\mu e_{\text{th}}\right)$ for given $e_{\text{th}}$. In other words, targeting an energy decay rate, $\mu$, we should have the constant energy arrival less than or equal to $u^{\star}$ such that we can keep the steady-state energy overflow probability less than or equal to $\exp\big(-\mu e_{\text{th}}\big)$. Moreover, it is important to note that while having the energy arrival below $u^{\star}$ decreases the energy overflow probability, it causes an increase in the energy outage probability.

\begin{prop}\label{pro:mu_zero_and_infinity_2}
When the energy decay rate, $\mu$, goes to zero, the maximum constant energy arrival approaches the average energy demand, $\mathbb{E}_{p}\left[p(i)\right]$. In particular,\[u^{\star}=-\lim_{\mu\to0}\lim_{t\to\infty}\frac{1}{t\mu}\ln\mathbb{E}_{p}\left[\exp\left(-\mu P(t)\right)\right]=\mathbb{E}_{p}\left[p(i)\right].\]Furthermore, when the energy decay rate, $\mu$, goes to infinity, the maximum constant energy arrival approaches the minimum energy demand, $\min\{p(i)\}$. In particular,\[u^{\star}=-\lim_{\mu\to\infty}\lim_{t\to\infty}\frac{1}{t\mu}\ln\mathbb{E}_{p}\left[\exp\left(-\mu P(t)\right)\right]=\min\{p(i)\}.\]
\end{prop}

\emph{Proof}: See Appendix \ref{app:proposition_mu_zero_and_infinity_2}.

\begin{rem}
Proposition \ref{pro:mu_zero_and_infinity} suggests that when the constant energy demand is less than or equal to the average value of the energy arrival process, energy overflows are inevitable. It also shows that, as long as the energy demand is greater than or equal to the maximum possible energy arrival in any time frame, there is no energy overflow in the battery. Likewise, according to Proposition \ref{pro:mu_zero_and_infinity_2}, when the value of a constant energy arrival process is greater than or equal to the average value of the energy demand process, the energy overflows are imminent, and when the constant energy arrival is less than or equal to the minimum energy demand, there is no energy overflow in the battery. Although the statements in Propositions \ref{pro:mu_zero_and_infinity} and \ref{pro:mu_zero_and_infinity_2} are rather intuitive, they are confirmed by the characterization in (\ref{eq:arrival_demand_balance}), where we express the probability that the energy level in the battery is above a defined threshold as an exponential function of the energy decay rate and the defined threshold. In other words, assuming that there is a constant energy demand while the energy arrival process has a stochastic nature, we can express the energy overflow probability with an exponential function as long as the constant energy demand is between the average and maximum values of the energy arrival process. Equivalently, given that there is a constant energy arrival process while the energy demand process is stochastic, we can express the energy overflow probability as an exponential function when the constant value of the energy arrival process is between the minimum and average values of the energy demand process.
\end{rem}

\subsection{Energy Outage}\label{sec:energy_outage}
Recall that when there is not enough energy in the battery, i.e., when the energy in the battery is less than what the transmitter demands, an \textit{energy outage} occurs; the transmitter consumes all the energy available in the battery, leaving the battery empty. We say that following an energy outage event, the battery enters state 0. Herein, we define a state space $\mathcal{W}=\{0,1,\cdots\}$ consisting of non-negative integers, where the state $w(i)$ at time $i$ denotes the consecutive number of times the harvested and stored energy successfully meet the energy demand since the last energy outage event. More specifically, assume that the $i^{\text{th}}$ time frame results in an energy outage, and we have $e(i)=0$; and hence, the battery is in state $w(i)=0$. In the subsequent $(i+1)^{\text{th}}$ time frame, if the harvested energy is greater than the demanded energy, i.e., $u(i+1)\geq p(i+1)$, the battery enters state $w(i+1)=1$, and stores the excess energy. However, if $u(i+1)<p(i+1)$, another energy outage occurs, and the battery remains in state 0, i.e., $w(i+1)=0$. Similarly, given that $w(i+1)=1$, the transition from state 1 to state 2 occurs when $u(i+1)+u(i+2)-p(i+1)\geq p(i+2)$. On the other hand, when $u(i+1)+u(i+2)-p(i+1)<p(i+2)$, the battery goes from state 1 to state 0. Similarly, given that $w(i+2)=2$, the battery transitions to state 3, i.e., $w(i+3)=3$, if $u(i+1)+u(i+2)+u(i+3)-p(i+1)-p(i+2)\geq p(i+3)$, and $w(i+3)=0$ otherwise. Generalizing the above observations and considering again an infinite-size battery, we can state that given $w(i+m-1)=m-1$, we have $w(i+m)=m$ if $\sum_{j=i+1}^{i+m}u(j)\geq\sum_{j=i+1}^{i+m}p(j)$, and $w(i+m)=0$ otherwise.

\begin{figure}
	\centering
	\includegraphics[width = 0.56\textwidth]{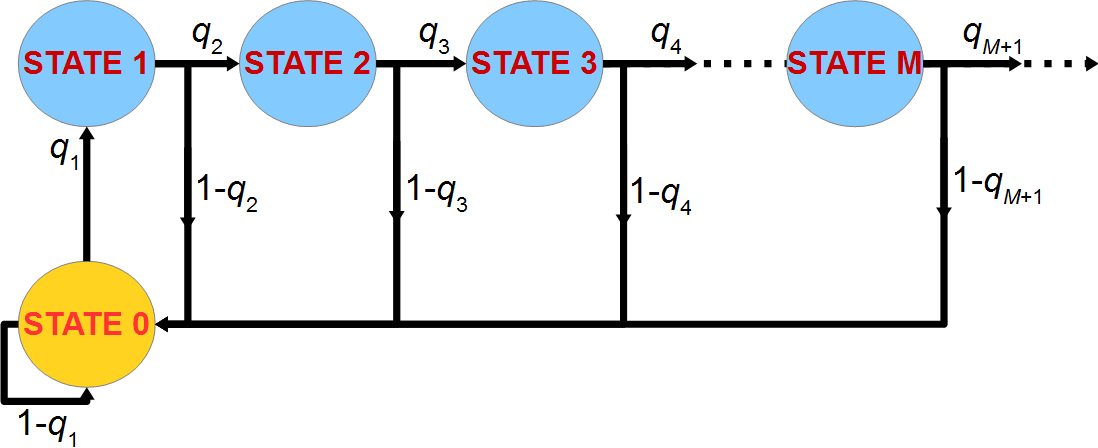}
	\caption{State transition model for the battery.} \label{fig:res_2}
		\vspace{-0.0cm}
\end{figure}

Notice that the battery in state $m-1$ for $m\in\{1,2,\cdots\}$ will go to either state $m$ or state 0. Given that the battery is in state $m-1$ in the $(i+m-1)^{\text{th}}$ time interval, we can express the state transition probability from state $m-1$ to state $m$ for $m\in\{1,2,\cdots\}$, denoted by $q_{m}(i+m)$, as
\begingroup
\allowdisplaybreaks
\begin{align}
&q_{m}(i+m)\triangleq\Pr\{w(i+m)=m|w(i+m-1)=m-1\}\label{eq:state_transition_probability_2_i}\\
&=\Pr\{w(i+m)=m|w(i+m-1)=m-1,w(i+m-2)=m-2,\cdots,w(i)=0\}\label{eq:state_transition_probability_2_f}\\
&=\frac{\Pr\{w(i+m)=m,\cdots,w(i)=0\}}{\Pr\{w(i+m-1)=m-1,\cdots,w(i)=0\}}\label{eq:state_transition_probability_2_e}\\
&=\frac{\Pr\left\{U(i+m,i)\geq P(i+m,i),\cdots,U(i+1,i)\geq P(i+1,i),e(i)=0\right\}}{\Pr\left\{U(i+m-1,i)\geq P(i+m-1,i),\cdots,U(i+1,i)\geq P(i+1,i),e(i)=0\right\}},\label{eq:state_transition_probability_2}
\end{align}
\endgroup
where $U(i+j,i)=U(i+j)-U(i)$ and $P(i+j,i)=P(i+j)-P(i)$ for $j\in\mathbb{Z}^{+}$. The transition probability from state $m-1$ to state 0 is $1-q_{m}(i+m)$. Above, (\ref{eq:state_transition_probability_2_f}) follows from the fact that the battery being in state $m-1$ in the $(i+m-1)^{\text{th}}$ frame has already been in states $0$ to $m-2$ in the time frames from the $i^{\text{th}}$ to the $(i+m-2)^{\text{th}}$. This also means that the battery was empty in the $i^{\text{th}}$ time frame, i.e., $e(i)=0$. In (\ref{eq:state_transition_probability_2_e}), we invoke Bayes' theorem. Moreover, since the energy arrival and demand processes are stochastic and ergodic, we can re-write (\ref{eq:state_transition_probability_2}) as
\begingroup
\allowdisplaybreaks
\begin{align}
q_{m}=\frac{\Pr\left\{U(m)\geq P(m),\cdots,U(1)\geq P(1)\right\}}{\Pr\left\{U(m-1)\geq P(m-1),\cdots,U(1)\geq P(1)\right\}}\label{eq:state_transition_probability_2_add}
\end{align}
\endgroup
for $m\in\{2,3,\cdots\}$, and $q_{1}=\Pr\left\{u(1)\geq p(1)\right\}$, where the battery is initially empty. Now, modeling the energy outages and consecutive energy supply guarantees as a Markov process, we have the state transition diagram given in Fig. \ref{fig:res_2}. Correspondingly, we can write the state transition matrix as follows:
\begin{align}\label{M}
M=\begin{pmatrix}
    1-q_1 & 1-q_2 & \cdots & 1-q_{m} & \cdots\\
    q_1 & 0 & \cdots & 0 & \cdots\\
    0 & q_2 & \cdots & 0 & \cdots\\
    \vdots &\vdots & \ddots&\vdots&\vdots\;\vdots\;\vdots\\
    0 & 0 & \cdots & q_{m} & \cdots\\
		\vdots &\vdots & \vdots\;\vdots\;\vdots&\vdots&\vdots\;\vdots\;\vdots
  \end{pmatrix}.
\end{align}
Now, let $\boldsymbol{\pi}$ be the vector of steady-state probabilities, i.e., $\boldsymbol{\pi}=\{\pi_{0},\pi_{1},\cdots,\pi_{m},\cdots\}^{T}$ satisfying $\boldsymbol{\pi}=M\boldsymbol{\pi}$ and $\sum_{i=0}^{\infty}\pi_{i}=1$, where $[\cdot]^{T}$ is the transpose operator. We can notice that the steady-state probability $\pi_{0}$ gives us the energy outage probability, which is $\pi_{0}=\frac{1}{1+\sum_{m=1}^{\infty}\prod_{i=1}^{m}q_{i}}$. Having $\pi_{0}$ and $M$, we can easily provide the other steady-state probabilities as follows: $\pi_{k}=\pi_{k-1}q_{k}=\pi_{0}\prod_{i=1}^{k}q_{i}=\frac{\prod_{i=1}^{k}q_{i}}{1+\sum_{m=1}^{\infty}\prod_{i=1}^{m}q_{i}}$ for $k\in\{1,2,\cdots\}$. Recalling that both $u(i)$ and $p(i)$ are stationary and ergodic, and that the stability condition given in Definition \ref{def:definition_decay_rate} is satisfied, we can show that $\pi_{0}\geq\pi_{1}\geq\pi_{2}\geq\cdots\geq\pi_{m}\geq\cdots$. Furthermore, using (\ref{eq:state_transition_probability_2_add}) and assuming that consecutive energy arrivals and consecutive energy demands are independent and identically distributed (i.i.d.), we can employ an inductive method and realize that $q_{1}\leq q_{2}\leq q_{3}\leq\cdots\leq q_{m}\leq\cdots$, and find an upper bound on the outage probability as follows:
\begingroup
\allowdisplaybreaks
\begin{align}
\pi_{0}&=\frac{1}{1+\sum_{m=1}^{\alpha-1}\prod_{i=1}^{m}q_{i}+\sum_{m=\alpha}^{\infty}\prod_{i=1}^{m}q_{i}}=\frac{1}{1+\sum_{m=1}^{\alpha-1}\prod_{i=1}^{m}q_{i}+\prod_{i=1}^{\alpha}q_{i}\left[1+\sum_{m=\alpha+1}^{\infty}\prod_{i=\alpha+1}^{m}q_{i}\right]}\nonumber\\
&\leq\frac{1}{1+\sum_{m=1}^{\alpha-1}\prod_{i=1}^{m}q_{i}+\frac{\prod_{i=1}^{\alpha}q_{i}}{1-q_{\alpha+1}}}\label{eq:upper_bound_pi_0}
\end{align}
\endgroup
for $\alpha\in\{1,2,\cdots\}$. The upper bound in (\ref{eq:upper_bound_pi_0}) comes from the fact that $\sum_{m=\alpha+1}^{\infty}\prod_{i=\alpha+1}^{m}q_{i}\geq\sum_{m=\alpha+1}^{\infty}\prod_{i=\alpha+1}^{m}q_{\alpha+1}$. With increasing $\alpha$, the bound becomes tighter. Performing the sum in the denominator in (\ref{eq:upper_bound_pi_0}) up to a finite number provides us the energy outage probability closely.

\section{Transmission Throughput}\label{sec:transmission_throughput}
In this section, we concentrate on the performance levels of the aforementioned EH wireless communication system. Therefore, we invoke the average data service rate in the wireless channel and the effective capacity in the data link layer as performance measures. The average data service rate is defined as the average number of bits the transmitter forwards to the receiver reliably. On the other hand, the effective capacity is the maximum constant data arrival rate at the transmitter buffer that the data service process can support under prescribed QoS constraints, i.e., data buffer overflow and buffering delay violation probability constraints. In the sequel, we provide expressions for the average data service rate and the effective capacity given a general class of energy arrival and demand processes, and then exemplify our results by constant energy demand rate and fixed data transmission rate for a clear presentation of our analysis.

\subsection{Average Data Service Rate}\label{sec:reliable_throughput}
We assume that the transmitter sets beforehand the energy demand process, $p(i)$, considering the energy overflow and outage probability constraints. Then, it adjusts the transmission strategy over the channel, $r(i)$, to maximize the average data service rate. As long as there is no energy outage, the transmitter does not change the transmission strategy. On the other hand, when there is an energy outage, the transmitter uses all the available energy in the battery and re-adjusts the transmission rate according to the available power, again to maximize the average data service rate\footnote{We assume that the transmitter is aware of the instantaneous battery state. Therefore, it is able to re-adjust its transmission rate according to the available energy in the battery.}. Although, the transmission rate policy is pre-determined along with the energy demand policy, it is indeed a function of the available energy in the battery given in (\ref{eq:power_level}), and hence it is stochastic. Another challenge in the wireless medium is the channel fading. When the transmission is exposed to deep fading in one frame, reliable decoding at the receiver may not be possible. In other words, if the instantaneous channel capacity falls below the transmission rate, a \emph{transmission outage} occurs; that is, the receiver fails to decode the data. Therefore, even if the desired energy rate is met, a transmission outage may occur due to channel fading.

We consider a block-fading channel model, and assume that the fading coefficient, $h(i)$, stays constant during one time frame and changes independently from one frame to the next. Now, given that the transmission power is as given in (\ref{eq:power_level}), the instantaneous channel capacity in the $i^{\text{th}}$ time frame is $\mathcal{I}(i)\triangleq N\log_{2}\left(1+\frac{|h(i)|^2p_c(i)}{N\sigma_{w}^{2}}\right)$ bits/frame, assuming Gaussian codebooks\footnote{One can easily adopt practical modulation techniques, e.g., binary phase-shift keying and quadrature amplitude modulations, into the framework provided in this paper. When practical modulation techniques are employed, one should consider the instantaneous mutual information between the channel input and output rather than the instantaneous channel capacity. We note that the instantaneous channel capacity is the maximum mutual information between the channel input and output.} are employed \cite[Ch. 9.1]{book_information_theory}\cite{caire1999optimum}. $N$ is the number of transmitted symbols in one time frame, which is assumed to be sufficiently large, and $\sigma_{w}^{2}$ is the variance of the zero-mean noise in the wireless channel. If $r(i)\leq \mathcal{I}(i)$, we assume that reliable transmission takes place and the receiver decodes the data successfully. Otherwise, a transmission outage occurs, and the effective transmission rate becomes zero. Recalling that $s(i)$ is the data service rate from the buffer in the $i^{\text{th}}$ time frame, we have: $s(i)=0$ if $r(i)>\mathcal{I}(i)$, and $s(i)=r(i)$ otherwise. Now, we can express the average data service rate in the steady-state as follows:
\begin{align}
\hspace{-0.5cm}s_{\text{avg}}=&\mathbb{E}_{u(i),p(i)}\left[s(i)\right]
=\mathbb{E}_{u(i),p(i)}\left[r(i)\cdot\textbf{1}\left[r(i)\leq\mathcal{I}(i)\right]\right]
=\mathbb{E}_{u(i),p(i)}\left[r(i)\cdot\textbf{1}\left[\kappa(i)\leq|h(i)|^{2}\right]\right],\label{eq:avg_srvc_rate_all}
\end{align}
where $\kappa(i)=\left(2^{\frac{r(i)}{N}}-1\right)\frac{N\sigma_{w}^{2}}{p_c(i)}$, and $\textbf{1}[x]$ is the indicator function, which is $1$ if $x$ is correct, and $0$ otherwise. Noting that the result in (\ref{eq:avg_srvc_rate_all}) is applicable in any setting with stationary and ergodic energy arrival and demand processes with finite means and variances, we consider the following specific case with constant energy demand and fixed data transmission rate in order to gain more insights.

\subsubsection{Constant energy demand rate and a fixed transmission rate}\label{subsubsec:constant_energy_rate}
Given a Rayleigh fading channel model\footnote{Although the result in (\ref{eq:avg_srvc_rate_all}) and the result in Theorem \ref{theo:effective_capacity} in the sequel are valid for channel models with arbitrary distributions with finite mean and variance, we focus on Rayleigh fading due to its practical relevance.}, i.e., the absolute power of the channel fading, $|h(i)|^2$, is exponentially distributed with parameter $\frac{1}{\sigma_{h}^{2}}$, where $\sigma_{h}^{2}$ is the variance of the channel fading, let us assume that we have an energy demand process with constant rate\footnote{In order to provide a smooth presentation of the aforementioned framework, we have considered the special case of constant energy demand rate as an example. However, one easily implement other energy demand policies. For instance, we have implemented and simulated a water-filling power allocation policy based energy demand policy in \cite{akin2017energy} under energy underflow probability constraint. Energy underflow refers to the case the energy level in the battery falls below a certain level.}, i.e., $p(i)=p$, and $\frac{p}{N}$ is the average symbol power. Hence, the consumed energy given in (\ref{eq:power_level}) is $p_{\text{c}}(i)= p$ if $p\leq\xi(i)$, and $p_{\text{c}}(i)=\xi(i)$ otherwise, where $\xi(i)=e(i-1)+u(i)$. Noting that the transmitter knows only the channel statistics but not its realizations, and that the transmitter sets the transmission rate according to the available energy, we assume that the data is transmitted at a fixed rate\footnote{Although our framework is good for data transmission settings with varying data traffic rate, we consider a fixed data transmission rate for mathematical tractability. We refer interested readers to \cite{pasch1993comparing} for more details on data traffic types. Voice and video traffic can be modeled with a constant data service rate.} as a function of the consumed energy, i.e., $r(i)=g(p_{c}(i))$ bits/frame, where $g(\cdot)$ is the pre-determined transmission rate function, which is monotonically increasing\footnote{We do not define a specific function but consider a general definition for the function that sets the transmission rate. In other words, $g(\cdot)$ is a matter of system design and may depend on the statistics of the energy demand policy. However, for instance, one can set $g(\cdot)$ as a linear function of the energy demand or the available energy in the battery.}. Then, we can express the average data service rate in the steady-state\footnote{After the battery reaches the steady-state, the probability of the battery being in state $m$ in any time frame becomes $\pi_{m}$, as defined in Section \ref{sec:energy_outage}, and it does not change with state transitions in the consecutive time frames.} as
\begingroup
\allowdisplaybreaks
\begin{align}\label{constant_transmission_rate_and_power}
s_{\text{avg}}=&\mathbb{E}_{\xi(i)}\left[r(i)\textbf{1}\left[r(i)\leq \mathcal{I}(i)\right]\right]=\int_{0}^{\infty}r(i)\Pr\left\{r(i)\leq \mathcal{I}(i)\right\}f_{\xi(i)}(\xi(i))d\xi(i)\\
=&\sum_{m=0}^{\infty}\int_{0}^{\infty}g(p_{c}(i))\Pr\left\{g(p_{c}(i))\leq \mathcal{I}(i)\right\}\Pr\{w(i-1)=m\}\nonumber\\
&\times f_{\xi(i)|w(i-1)=m}(\xi(i)|w(i-1)=m)d\xi(i)\label{eq:constant_transmission_rate_3}\\
=&\sum_{m=0}^{\infty}\pi_{m}\int_{0}^{\infty}g(p_{c}(i))\Pr\left\{g(p_{c}(i))\leq \mathcal{I}(i)\right\}f_{\xi(i)|w(i-1)=m}(\xi(i)|w(i-1)=m)d\xi(i)\\
=&\sum_{m=0}^{\infty}\pi_{m}\int_{0}^{p}g(\xi(i))\Pr\left\{g(\xi(i))\leq \mathcal{I}(i)\right\}f_{\xi(i)|w(i-1)=m}(\xi(i)|w(i-1)=m)d\xi(i)\nonumber\\
&+\sum_{m=0}^{\infty}\pi_{m}\int_{p}^{\infty}g(p)\Pr\left\{g(p)\leq \mathcal{I}(i)\right\}f_{\xi(i)|w(i-1)=m}(\xi(i)|w(i-1)=m)d\xi(i)\label{eq:constant_transmission_rate_5}\\
=&\sum_{m=0}^{\infty}\pi_{m}\int_{0}^{p}g(\xi(i))\Pr\left\{\kappa(\xi(i))\leq|h(i)|^{2}\right\}f_{\xi(i)|w(i-1)=m}(\xi(i)|w(i-1)=m)d\xi(i)\nonumber\\
&+\sum_{m=0}^{\infty}\pi_{m}\int_{p}^{\infty}g(p)\Pr\left\{\kappa(p)\leq|h(i)|^2\right\}f_{\xi(i)|w(i-1)=m}(\xi(i)|w(i-1)=m)d\xi(i)\\
=&\sum_{m=0}^{\infty}\pi_{m}\int_{0}^{p}g(\xi(i))e^{\frac{-\kappa(\xi(i))}{\sigma_{h}^{2}}}f_{\xi(i)|w(i-1)=m}(\xi(i)|w(i-1)=m)d\xi(i)\nonumber\\
&+\sum_{m=0}^{\infty}\pi_{m}g(p)e^{\frac{-\kappa(p)}{\sigma_{h}^{2}}}\int_{p}^{\infty}f_{\xi(i)|w(i-1)=m}(\xi(i)|w(i-1)=m)d\xi(i)\\
=&\sum_{m=0}^{\infty}\pi_{m}\int_{0}^{p}g(\xi(i))e^{\frac{-\kappa(\xi(i))}{\sigma_{h}^{2}}}f_{\xi(i)|w(i-1)=m}(\xi(i)|w(i-1)=m)d\xi(i)\nonumber\\
&+g(p)e^{\frac{-\kappa(p)}{\sigma_{h}^{2}}}\sum_{m=0}^{\infty}\pi_{m}q_{m+1}\\
=&\sum_{m=0}^{\infty}\pi_{m}\int_{0}^{p}g(\xi(i))e^{\frac{-\kappa(\xi(i))}{\sigma_{h}^{2}}}f_{\xi(i)|w(i-1)=m}(\xi(i)|w(i-1)=m)d\xi(i)\nonumber\\
&+g(p)e^{\frac{-\kappa(p)}{\sigma_{h}^{2}}}(1-\pi_{0}),\label{eq:avg_throughput_final}
\end{align}
\endgroup
where $f_{\xi(i)}\left(\xi(i)\right)$ and $f_{\xi(i)|w(i-1)=m}\left(\xi(i)|w(i-1)=m\right)$ are the probability density function of $\xi(i)$, and the conditional probability density function of $\xi(i)$ in the $i^{\text{th}}$ time frame, respectively. In (\ref{eq:constant_transmission_rate_5}), $g(p)$ is the fixed transmission rate when the battery satisfies the constant energy demand rate, and $g(\xi(i))$ is the re-adjusted data transmission rate according to the available energy in the battery when the battery cannot sustain the constant energy demand rate. Moreover, we define $\kappa(a)\triangleq\left(2^{\frac{g(a)}{N}}-1\right)\frac{N\sigma_{w}^{2}}{a}$ for $a\in{\mathbb{R}^{+}}$. In (\ref{eq:constant_transmission_rate_3}), $\pi_{m}\triangleq\Pr\{w(i-1)=m\}$ is the probability that the battery is in state $m$ in the $(i-1)^{\text{th}}$ time frame in the steady-state.

We obtain the average data service rate in (\ref{eq:avg_throughput_final}) for an arbitrary transmission rate; and therefore, it can be maximized with the appropriate choice of the transmission rate. Particularly, we have the maximized average data service rate as follows:
\begin{align}\label{eq:s_avg_max_ad}
s_{\text{avg}}^{\max}=&\max_{g(p)}\left\{g(p)e^{\frac{-\kappa(p)}{\sigma_{h}^{2}}}\right\}(1-\pi_{0})\nonumber\\
&+\sum_{m=0}^{\infty}\pi_{m}\int_{0}^{p}\max_{g(\xi(i))}\left\{g(\xi(i))e^{\frac{-\kappa(\xi(i))}{\sigma_{h}^{2}}}\right\}f_{\xi(i)|w(i-1)=m}(\xi(i)|w(i-1)=m)d\xi(i).
\end{align}

Notice that (\ref{eq:avg_throughput_final}) is a sum of infinite number of terms. Therefore, it may be difficult obtain a closed-form solution. However, we can lower-bound the expression in (\ref{eq:avg_throughput_final}) for any $\alpha>0$ as follows:
\begingroup
\allowdisplaybreaks
\begin{align}
s_{\text{avg}}=&g(p)e^{\frac{-\kappa(p)}{\sigma_{h}^{2}}}(1-\pi_{0})+\sum_{m=0}^{\infty}\pi_{m}\int_{0}^{p}g(\xi(i))e^{\frac{-\kappa(\xi(i))}{\sigma_{h}^{2}}}f_{\xi(i)|w(i-1)=m}(\xi(i)|w(i-1)=m)d\xi(i)\nonumber\\
=&g(p)e^{\frac{-\kappa(p)}{\sigma_{h}^{2}}}(1-\pi_{0})+\sum_{m=0}^{\alpha}\pi_{m}\int_{0}^{p}g(\xi(i))e^{\frac{-\kappa(\xi(i))}{\sigma_{h}^{2}}}f_{\xi(i)|w(i-1)=m}(\xi(i)|w(i-1)=m)d\xi(i)\nonumber\\
&+\sum_{m=\alpha+1}^{\infty}\pi_{m}\int_{0}^{p}g(\xi(i))e^{\frac{-\kappa(\xi(i))}{\sigma_{h}^{2}}}f_{\xi(i)|w(i-1)=m}(\xi(i)|w(i-1)=m)d\xi(i)\nonumber\\
\geq&g(p)e^{\frac{-\kappa(p)}{\sigma_{h}^{2}}}(1-\pi_{0})+\sum_{m=0}^{\alpha}\pi_{m}\int_{0}^{p}g(\xi(i))e^{\frac{-\kappa(\xi(i))}{\sigma_{h}^{2}}}f_{\xi(i)|w(i-1)=m}(\xi(i)|w(i-1)=m)d\xi(i)\label{eq:lower_bound_avg_rate}
\end{align}
\endgroup
The lower bound in (\ref{eq:lower_bound_avg_rate}) converges to the actual average data service rate with increasing $\alpha$ because the steady-state probability, $\pi_{i}$, decreases with the increasing battery state index. Therefore, we can approximate the average data service rate by taking the sum up to a finite value. In our simulations, we observe that after $\alpha=100$, the average data service rate values do not change. Herein, one can find the data transfer rate function, $g(\cdot)$, that maximizes the lower bound similarly as in (\ref{eq:s_avg_max_ad}).

\subsection{Effective Capacity}\label{sec:effective_capacity}
Recall that we store the data arriving at the transmitter in the data buffer before sending it in frames of $N$ symbols. Therefore, buffer overflow and delay bounds can be addressed by imposing statistical constraints on the queue length and delay in the buffer. Thus, we set the effective capacity as the performance measure in order to take into account the data queueing constraints of the aforementioned delay-limited EH system. We can define the effective capacity as the maximum constant data arrival rate that a given stochastic service process can support in order to satisfy the desired QoS requirements specified by the QoS exponent $\theta$. We can formulate the effective capacity as \cite{wu_negi}
\begin{align}\label{effective_capacity}
C_{E}(\theta)=-\lim_{t\to\infty}\frac{1}{t\theta}\ln\mathbb{E}\left[\exp\left(-\theta S(t)\right)\right],
\end{align}
where $S(t)\triangleq\sum_{i=1}^{t}s(i)$ is the time-accumulated service process, and $s(i)$ is the discrete-time stationary and ergodic data service process. Notice that $\lim_{t\to\infty}\frac{1}{t}\ln\mathbb{E}\left[\exp\left(\theta S(t)\right)\right]$ is the asymptotic log-moment generating function of $S(t)$. Further notice that when $\theta$ goes to zero in limit, i.e., when there are no QoS constraints, the effective capacity in (\ref{effective_capacity}) converges to the average data service rate.

We express the QoS exponent, which describes the decay rate of the tail distribution of the queue length, $d(i)$, as
\begin{equation}\label{theta_p}
\theta=-\lim_{d_{\text{th}}\to\infty}\frac{\ln\Pr\{d\geq d_{\text{th}}\}}{d_{\text{th}}},
\end{equation}
where $d$ denotes the steady-state queue length, and $\Pr\{d\geq d_{\text{th}}\}$ is the data buffer overflow probability for a given threshold $d_{\text{th}}$. Similarly to the approximation of the energy overflow probability, we can have an exponential approximation also for the data buffer overflow probability when we have a large data buffer size as follows: $\Pr\{d\geq d_{\text{th}}\}\approx \exp(-\theta d_{\text{th}})$. We can now easily infer that larger $\theta$ describes stricter QoS requirements because the data buffer overflow probability for a given threshold decreases with increasing $\theta$, while smaller $\theta$ indicates less strict QoS constraints due to the fact that the data buffer overflow probability increases with decreasing $\theta$.

As seen in (\ref{effective_capacity}), the effective capacity depends on the data service process, $s(i)$, and hence, the data transmission rate in the channel, $r(i)$. Because the transmission rate is a function of the consumed energy, it is also a function of the energy demand and the energy level in the battery. Therefore, we can easily see that the effective capacity, which is the maximum data arrival rate that the service can support under QoS constraints, is affected by the energy arrival and demand processes. In the following, we provide the effective capacity of the aforementioned delay-limited wireless EH system.
\begin{theo}\label{theo:effective_capacity}
{The effective capacity of the point-to-point wireless communication system in which the transmitter performs EH, stores the harvested energy in a battery, and is subject to a given data QoS exponent, $\theta$, and an energy decay rate, $\mu$, is given as
\begin{equation}\label{effective_capacity_main}
C_{E}(\theta,\mu)=-\frac{1}{N\theta}\ln\left(\chi^{\star}\right)\text{ bits/channel use,}
\end{equation}
where $\chi^{\star}$ is the unique real positive root of $z(\chi)$, which is defined as
\begin{align}\label{characteristic_function}
z(\chi)=&\lim_{m\to\infty}\Bigg[\chi^{m}-\phi_{0}(-\theta)\sum_{n=1}^{m}\chi^{m-n}(1-q_{n})\prod_{j=1}^{n-1}q_{j}\phi_{j}(-\theta)\Bigg],
\end{align}
where
\begin{align}
\phi_{0}(\theta)=&\frac{1}{\pi_{0}}\sum_{m=0}^{\infty}\pi_{m}\mathbb{E}_{p(i)}\Bigg[\int_{0}^{p(i)}\Big[\exp\left(\theta g(\xi(i))\right)\Pr\left\{\kappa(\xi(i))\leq|h(i)|^{2}\right\}+\Pr\left\{\kappa(\xi(i))>|h(i)|^{2}\right\}\Big]\nonumber\\
&\times f_{\xi(i)|w(i-1)=m}(\xi(i)|w(i-1)=m)d\xi(i)\Bigg]\label{moment_generating_function_zero_final}
\end{align}
and
\begin{align}
\phi_{j}(\theta)=&\frac{1}{q_{j}}\mathbb{E}_{p(i)}\Bigg[\int_{p(i)}^{\infty}\Big[\exp\left(\theta g(p(i))\right)\Pr\left\{\kappa(p(i))\leq|h(i)|^2\right\}+\Pr\left\{\kappa(p(i))>|h(i)|^2\right\}\Big]\nonumber\\
&\times f_{\xi(i)|w(i-1)=j-1}(\xi(i)|w(i-1)=j-1)d\xi(i)\Bigg]\label{moment_generating_function_j_final}
\end{align}
for $j\in\{1,\cdots\}$, where the buffer size, $d_{\text{th}}$, is large. Recall that $p_{c}(i)=\xi(i)=e(i-1)+u(i)$ when $e(i-1)+u(i)<p(i)$ and $p_{c}(i)=p(i)$ when $e(i-1)+u(i)\geq p(i)$. Moreover, an upper bound to the effective capacity is given by $C_{E}(\theta,\mu)=-\frac{1}{N\theta}\ln\left(\chi^{\star}\right)\leq-\frac{1}{N\theta}\ln\left(\chi_{\alpha}^{\star}\right)$, where $\chi_{\alpha}^{\star}$ is the unique real positive root of $z^{\star}(\chi_{\alpha})$, defined as
\begin{align}\label{characteristic_function_upper_bound}
z^{\star}(\chi_{\alpha})=&(\chi_{\alpha})^{\alpha}-\phi_{0}(-\theta)\sum_{n=1}^{\alpha}(\chi_{\alpha})^{\alpha-n}(1-q_{n})\prod_{j=1}^{n-1}q_{j}\phi_{j}(-\theta).
\end{align}
The aforementioned upper bound converges to the effective capacity as $\alpha$ is increased.}
\end{theo}

\emph{Proof}: See Appendix \ref{app:theorem_effective_capacity}.

\subsubsection{Constant energy demand rate and fixed transmission rate}
Let us consider the channel and transmission settings described in Section \ref{subsubsec:constant_energy_rate}. Then, we can express the moment generating functions in (\ref{moment_generating_function_zero_final}) and (\ref{moment_generating_function_j_final}) as follows:
\begin{align*}
\phi_{0}(\theta)=&\frac{1}{\pi_{0}}\sum_{m=0}^{\infty}\pi_{m}\int_{0}^{p}\bigg[\exp\left(\theta g(\xi(i))\right)\exp\left(-\frac{\kappa(\xi(i))}{\sigma_{h}^{2}}\right)+1-\exp\left(-\frac{\kappa(\xi(i))}{\sigma_{h}^{2}}\right)\bigg]\\
&\times f_{\xi(i)|w(i-1)=m}(\xi(i)|w(i-1)=m)d\xi(i),
\end{align*}
and $\phi_{j}(\theta)=\exp\left(\theta g(p)\right)\exp\left(-\frac{\kappa(p)}{\sigma_{h}^{2}}\right)+1-\exp\left(-\frac{\kappa(p)}{\sigma_{h}^{2}}\right)$, respectively.

\section{Numerical Results}\label{sec:numerical_results}
We substantiate our theoretical results with numerical demonstrations. In the following, we initially show results regarding the battery constraints, i.e., the energy overflow and outage probabilities. Particularly, we compare the theoretical approximations with finite-size and infinite-size battery simulations. Then, we plot the average data service rate as a function of the energy decay rate, $\mu$. Finally, we have the effective capacity results under QoS constraints with different energy arrival processes.

\subsection{Energy Overflow and Outage Probabilities}
\begin{figure*}[t]
	\centering
	\subfigure[Energy overflow probability, $\Pr\{e\geq e_{\text{th}}\}$, vs. energy threshold, $e_{\text{th}}$.]{
	\includegraphics[width=\figsize\textwidth]{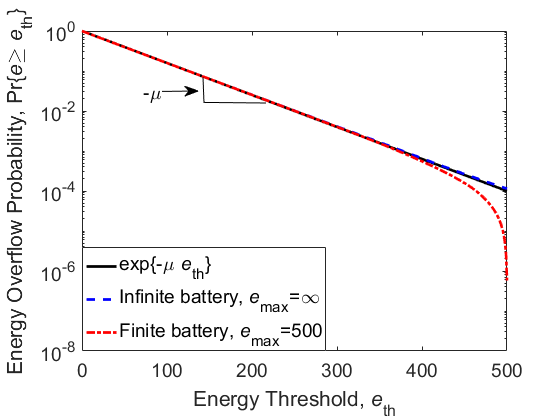}\label{fig:fig_1}}
	\quad
	\subfigure[Energy outage probability, $\pi_{0}$, vs. energy decay rate, $\mu$ (dB).]{
	\includegraphics[width=\figsize\textwidth]{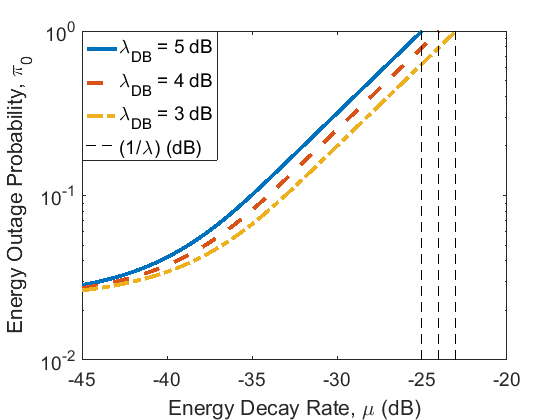}\label{fig:fig_2}}
	\caption{Energy overflow and energy outage probabilities.}
	\vspace{-0.0cm}
\end{figure*}

We model the energy arrival process using a Weibull distribution, which is widely employed in the wind industry as the preferred approach for modeling the wind speed for energy assessment due to its high versatility, flexibility, and accuracy for describing the wind speed variations \cite{olaofe2012statistical}. Hence, we have the following probability density function for the energy arrival samples for $u\geq0$: $f_{u}(u)=\frac{k}{\lambda}\left(\frac{u}{\lambda}\right)^{k-1}\exp\left(-\left(\frac{u}{\lambda}\right)^{k}\right)$, where $k$ and $\lambda$ are shape and scale parameters, respectively. We also note that we employ i.i.d. energy samples. In Fig. \ref{fig:fig_1}, we set the desired energy overflow probability to $\Pr\{e\geq e_{\max}\}=10^{-4}$ for a threshold value of $e_{\max}=500$ energy units. Then, using the exponential approximation, i.e., $\Pr\{e\geq e_{\text{th}}\}\approx\exp(-\mu e_{\text{th}})$, we find $\mu=0.0184$. We initially plot the energy overflow probability using the exponential approximation, and then compare it with the simulation results where we obtain the energy overflow probability with an infinite-size battery and a finite-size battery with $e_{\max}=500$ units. In the simulations, we have $k=5$ and $\lambda=2$ for the shape and scale parameters, respectively. We further run an energy demand process with constant rate that provides the equality in (\ref{eq:arrival_demand_balance}) for desired $\mu$ and is greater than the average energy arrival rate in order to guarantee the stability of the battery. However, one can implement the simulations with varying energy demand rates as long as (\ref{eq:arrival_demand_balance}) is provided and the stability condition, i.e., $\mathbb{E}_{u}\left[u(i)\right]<\mathbb{E}_{p}\left[p(i)\right]$, is guaranteed. The energy overflow probability approximation captures the simulation performances with the infinite-size and finite-size batteries very closely, while the energy overflow probability with the finite-size battery is less than the approximation for threshold values close to the battery size. However, we can accurately approximate the energy overflow probability for threshold values less than the battery size. Our approximation matches the finite-size battery simulations very closely for energy threshold values up to $80\%$ of the battery capacity, which is $400$ energy units in our simulations. In a real setting, one should not charge a battery completely but up to $80\%$ in order to improve the battery life-span and the energy efficiency \cite{battery_uni_2,website_2}.

In Fig. \ref{fig:fig_2}, we plot the energy outage probability, $\pi_{0}$ in (\ref{eq:upper_bound_pi_0}), as a function of the energy decay rate, $\mu$, for different scale parameters, $\lambda_{\text{dB}}=10\log_{10}\frac{\lambda}{N\sigma_{w}^{2}}=5$, $4$ and $3$ dB, when the shape parameter is $k=1$. Note that the mean value of the Weibull distribution is $\lambda\Gamma(1+1/k)$, where $\Gamma(\cdot)$ is the Gamma function. So, $\lambda$ becomes the average energy arrival rate when $k=1$. If we use all the energy for data transmission, we can consider $\lambda_{\text{dB}}$ as the average signal-to-noise ratio in the channel. In addition, given the energy decay rate, we determine the constant energy demand rate. Particularly, given that the energy packets arriving at the battery are i.i.d., the constant energy demand rate is $p=\frac{1}{\mu}\ln\mathbb{E}_{u}\left[e^{\mu u}\right]=\frac{1}{\mu}\ln\left(\frac{1}{1-\lambda\mu}\right)$ for $0<\mu<\frac{1}{\lambda}$. Notice that when $\mu$ goes to zero, $p$ approaches the average energy arrival rate, $\lambda$, whereas when $\mu$ goes to $\frac{1}{\lambda}$, $p$ approaches infinity. Therefore, the outage probability becomes 1 once $\mu$ is greater than $\frac{1}{\lambda}$ for a given energy arrival process with the aforementioned distribution, which is displayed in Fig. \ref{fig:fig_2}. We can also infer that the energy outage probability goes to 1 when we consume the energy in the battery faster, i.e., as $\mu$ increases. The black dashed vertical lines in Fig. \ref{fig:fig_2} indicate the energy decay rate above which the constant energy demand rate is infinite for the defined energy arrival process, i.e., $p=\infty$. Therefore, when the energy storage conditions are stricter, energy demand processes without a constant rate should be favored. However, having a transmission system with varying transmission power capability leads to complexity in the transmitter design and a large number of calculation steps during data transmission. Furthermore, having an energy decay rate (or an energy overflow probability for a fixed battery threshold), we can infer from the results that with the increasing scale parameter in the energy arrival process, the energy outage probability increases, which is due to the increased scattering in the probability distribution of the energy arrival samples with the increasing scale parameter.
\begin{figure*}[t]
	\centering
	\subfigure[AWGN channel.]{
	\includegraphics[width=\figsize\textwidth]{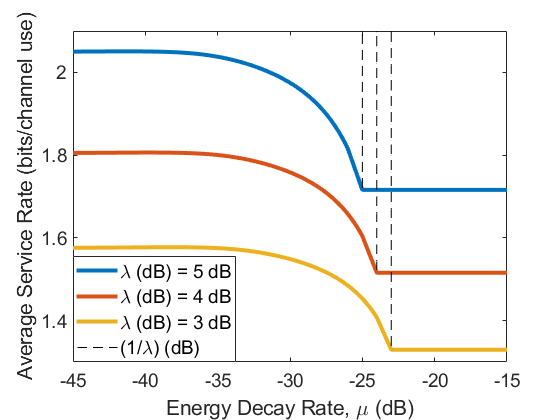}\label{fig:fig_3}}
	\quad
	\subfigure[Rayleigh channel.]{
	\includegraphics[width=\figsize\textwidth]{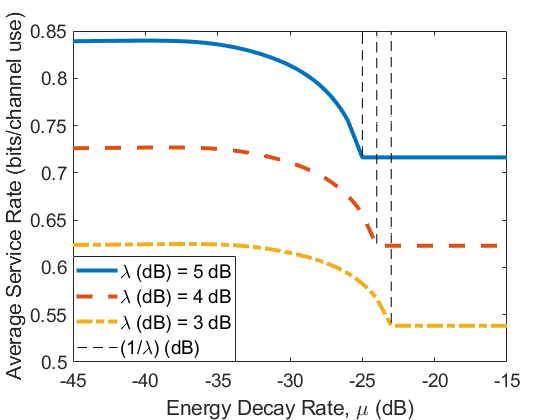}\label{fig:fig_4}}
	\caption{Average data service rate, $s_{\text{avg}}$, vs. energy decay rate, $\mu$, with different average energy arrivals, i.e., $\lambda_{\text{dB}}=5$, $4$ and $3$ dB.}
	\vspace{-0.0cm}
\end{figure*}

\subsection{Average Data Service Rate}
In Fig. \ref{fig:fig_3} and Fig. \ref{fig:fig_4}, we plot the average data service rate, $s_{\text{avg}}$, as a function of the energy decay rate, $\mu$, in AWGN and Rayleigh fading channels, respectively, given that we have an energy demand process with constant rate, and a constant transmission rate policy. In the case of an AWGN channel, we have a constant, unit-valued channel gain. Since the channel does not change, we set the transmission rate equal to the instantaneous mutual information, i.e., $r(i)=g(p)=N\log_{2}\Big(1+\frac{p}{N\sigma_{w}^{2}}\Big)$, when the power demand is satisfied. On the other hand, the transmission rate is set to $r(i)=g(\xi(i))=N\log_{2}\Big(1+\frac{\xi(i)}{N\sigma_{w}^{2}}\Big)$ when there is a power outage. Notice that $s(i)$ is always equal to $r(i)$, i.e., $s(i)=r(i)$, because there is no transmission outage since $r(i)=\mathcal{I}(i)$. When we have Rayleigh fading with unit variance, i.e., $\sigma_{h}^{2}=1$, we choose the transmission rate in a way to maximize the average data service rate in the corresponding time frame, i.e., $r(i)=\arg\max_{g(p)}\{g(p)\exp(\kappa(p))\}$ when the battery sustains the energy demand or $r(i)=\arg\max_{g(\xi(i))}\{g(\xi(i))\exp(\kappa(\xi(i)))\}$ when there is an energy outage, where $\kappa(a) = \left(2^{\frac{g(a)}{N}}-1\right)\frac{N\sigma_{w}^{2}}{\xi(i)}$ for $a\geq0$. Note that $\exp(\kappa(a))$ is the probability that the data transmission is successful in the corresponding time frame. The time frames are equal to 100 channel uses, i.e., $N=100$. For both channels, we again obtain results\footnote{Here, we simulate the energy arrivals and the channel fading gains with respect to their distributions assuming that Gaussian codebooks are employed. However, one can do simulations considering practical modulation techniques and their corresponding instantaneous mutual information values given the channel conditions.} with different scale parameters. We observe that increasing battery overflow constraints above certain values causes a sharp decrease in the average data service rate. Noting that the black dashed vertical lines indicate the energy decay rate above which the constant energy demand rate is infinite for the defined energy arrival process, i.e., the transmitter consumes the energy packets in the time frame they arrive at the battery when $\mu\geq\frac{1}{\lambda}$, the average data service rate becomes constant for $\mu\geq\frac{1}{\lambda}$.

\subsection{Effective Capacity}
\begin{figure*}[t]
	\centering
	\subfigure[AWGN channel.]{
	\includegraphics[width=\figsize\textwidth]{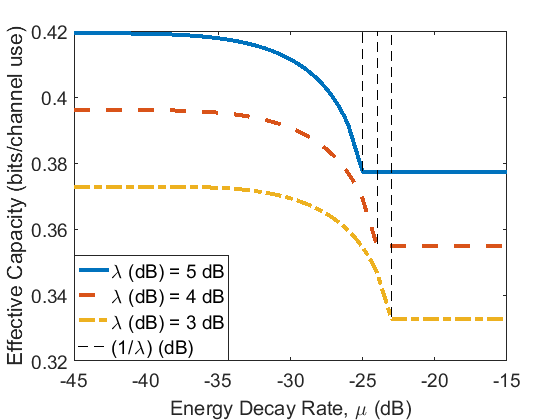}\label{fig:fig_5}}
	\quad
	\subfigure[Rayleigh channel.]{
	\includegraphics[width=\figsize\textwidth]{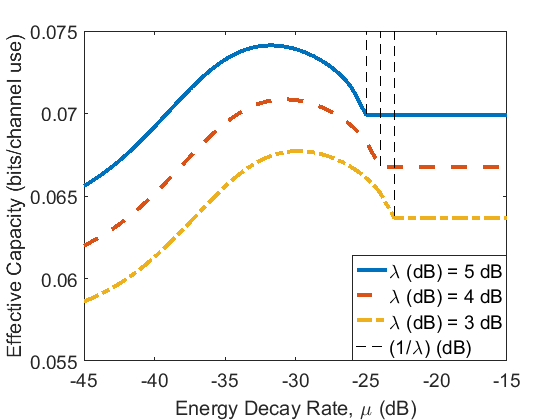}\label{fig:fig_7}}
	\caption{Effective capacity vs. energy decay rate, $\mu$, with different average energy arrivals, i.e., $\lambda_{\text{dB}}$, $4$ and $3$ dB.}
	\vspace{-0.0cm}
\end{figure*}

\begin{figure*}[t]
	\centering
	\subfigure[AWGN channel.]{
	\includegraphics[width=\figsize\textwidth]{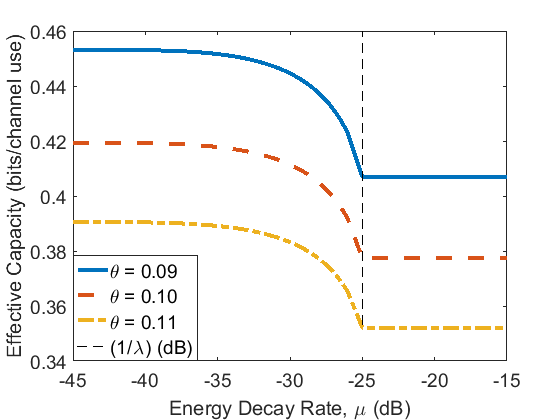}\label{fig:fig_6}}
	\quad
	\subfigure[Rayleigh channel.]{
	\includegraphics[width=\figsize\textwidth]{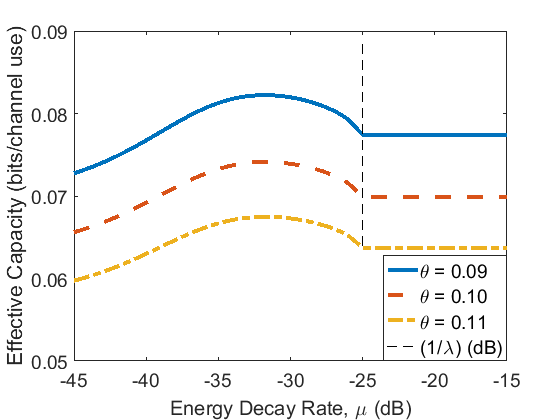}\label{fig:fig_8}}
	\caption{Effective capacity vs. energy decay rate, $\mu$, with different data transmission delay constraints, i.e., $\theta_{p}$.}
	\vspace{-0.0cm}
\end{figure*}

In Fig. \ref{fig:fig_5} and Fig. \ref{fig:fig_7}, we plot the effective capacity, $C_{E}(\theta,\mu)$, versus the energy decay rate, $\mu$, in AWGN and Rayleigh fading channels, respectively, again for energy processes with different average values. We further set the QoS exponent to $\theta=0.1$. Similar to the results in Fig. \ref{fig:fig_3}, the effective capacity of AWGN channels increases with decreasing $\mu$ as seen in Fig. \ref{fig:fig_5}, whereas unlike the results in Fig. \ref{fig:fig_4}, the effective capacity of Rayleigh fading channels rises with $\mu$ up to a certain value, and then it starts decreasing until $\mu$ reaches $\frac{1}{\lambda}$ as seen in Fig. \ref{fig:fig_7}. The only concern in AWGN channel is the energy outages, and the frequency of energy outages decreases with the decreasing energy demand rate, i.e., decreasing $\mu$. The effective capacity increases with decreasing $\mu$ in AWGN channel because the service process becomes more deterministic. On the other hand, there are two concerns in Rayleigh fading channel, namely, transmission outages and energy outages. While the frequency of energy outages decreases with decreasing $\mu$, the occurrence of transmission outages increases with decreasing $\mu$, and hence the effective capacity decreases. With increasing $\mu$, the energy outages become dominant; as a result, the transmitter cannot take advantage of higher channel fading gains when there is possibly very little energy in the battery. Therefore, the effective capacity first increases and then decreases with increasing $\mu$. Recall that when $\mu\geq\frac{1}{\lambda}$, the transmitter utilizes the energy packets as soon as they arrive at the battery. Specifically, when the buffer overflow concerns are of importance in data transmission with an energy demand process with constant rate, it is strategic to set the average transmission power to the average energy arrival rate in AWGN channels, while it is necessary to set to a value that is greater than the average energy arrival rate in Rayleigh fading channels. Subsequently in Fig. \ref{fig:fig_6} and Fig. \ref{fig:fig_8}, keeping the average energy arrival rate fixed at $\lambda_{\text{dB}}=5$ dB, we plot the effective capacity versus the energy decay rate for different QoS exponents, e.g., $\theta=0.09$, $0.10$ and $0.11$.
We can see that the QoS exponent does not impact the range of the energy decay rate, while it affects the performance levels substantially. We can remark that if the battery conditions are not very strict, we can increase the effective capacity by minimizing the energy decay rate in AWGN channels even though it will cause an increase in energy overflow probability. In Rayleigh fading channels, by setting the energy decay rate to a value between zero and $\frac{1}{\lambda}$, we can increase the effective capacity even though the average data service rate in Rayleigh fading channels is maximized when the energy decay rate is minimized as seen in Fig. \ref{fig:fig_8}. Depending on the transmission objective, a system designer can opt for an energy consumption and battery sustaining policy. One additional note is that when we compare the results in Fig. \ref{fig:fig_3} and Fig. \ref{fig:fig_4} with the results in Fig. \ref{fig:fig_5} and Fig. \ref{fig:fig_7}, we see a significant difference in performance levels of the average data service rate and the effective capacity. In other words, the service rate in the wireless channel from the transmitter to the receiver is much greater than the data arrival rate at the transmitter. This is because of the scattering in the data service rates that follows due to the energy outages and the transmission outages. More specifically, if there are no random energy outages and random transmission outages, the data arrival rate at the transmitter buffer will be equal to the constant transmission rate in the channel. Moreover, if we are to operate under data buffering constraints in Rayleigh channels, we need to have a moderate energy demand process. Otherwise, we can maximize the average data service rate by decreasing $\mu$.

\begin{rem} Different from the existing effective capacity studies in wireless fading channels, for instance see \cite{tang,musavian2010effective1,akin_eura,musavian2010effective,akin2010effective,elalem}, that invoke average and/or peak average power constraints, we do not have an energy source that provides guaranteed energy levels when needed, and due to the stochastic nature of the energy source, we rather employ the energy overflow and outage probability constraints in the battery. We further provide the effective capacity performance analysis as a function of the energy decay rate but not the signal-to-noise ratio since the energy decay rate is a tool that captures the balance between the energy overflow and outage probabilities, i.e., smaller energy decay rate implies increased energy overflow probability and decreased energy outage probability whereas higher energy decay rate indicates decreased energy overflow probability and increased energy outage probability.
\end{rem}

\vspace{-0.0cm}
\section{Conclusion}\label{sec:conclusion}
We have considered an energy harvesting transmitter equipped with a rechargeable battery and a data buffer. We have assumed that the transmitter harvests energy in random amounts and stores it in its battery, while the data also arrives at the data buffer in a random manner. We have provided a methodology to derive the relationship between the energy arrival and energy demand processes, and its impact on the throughput performance. We have initially approximated the energy overflow probability in the battery as an exponential function using tools from large deviation theory. Subsequently, projecting the energy availability at the battery on a Markov process, we have obtained the energy outage probability in the battery. Then, under the energy overflow and outage constraints, we have characterized the average data service rate over the wireless channel and the effective capacity in the data buffer under QoS constraints. We have substantiated our analytical results by numerical simulations considering AWGN and Rayleigh fading channels. Our results show that a strategy that stores the harvested energy and utilizes it regarding energy overflow and outage constraints results in better performance levels when compared to a strategy that consumes energy as soon as it is harvested. Finally, our results reveal that a strategy that saves energy as much as possible and avoids energy outages completely does not yield the maximum throughput performance under QoS constraints in Rayleigh fading channels. However, a strategy that consumes energy neither moderately nor greedily results in the maximum effective capacity performance.

\appendix

\subsection{Proof of Proposition \ref{pro:mu_zero_and_infinity}}\label{app:proposition_mu_zero_and_infinity}
When the energy decay rate, $\mu$, goes to zero, the minimum constant energy demand rate goes to the average energy arrival rate. In particular,
\begingroup
\allowdisplaybreaks
\begin{align}
p^{\star}&=\lim_{\mu\to0}\frac{1}{\mu}\Lambda_{u}(\mu)=\lim_{\mu\to0}\lim_{t\to\infty}\frac{1}{\mu t}\ln\mathbb{E}_{u}\Big[\exp\Big(\mu\sum_{i=1}^{t}u(i)\Big)\Big]\nonumber\\
&=\mathbb{E}_{u}\big[u(i)\big]+\lim_{\mu\to0}\frac{\mu}{2t_{k}}\sigma_{t_{k}}^{2}=\mathbb{E}_{u}\big[u(i)\big],
\end{align}
\endgroup
where $t_{k}$ is a large positive integer and $\sigma_{t_{k}}^{2}$ is the variance of $\sum_{i=1}^{t_{k}}u(i)$ \cite[Section III]{soret2010capacity}.

When the decay rate, $\mu$, goes to infinity, the minimum constant energy demand rate approaches the maximum energy arrival rate, i.e., $p^{\star}=\max\{u(i)\}=u_{\max}$. Noting that the limit when $\mu$ goes to infinity exists uniformly for any given $t>0$, we can again show this by exchanging the limits by invoking \cite[Theorem 1]{kadelburg2005interchanging}. Specifically,
\begingroup
\allowdisplaybreaks
\begin{align}
	p^{\star}&=\lim_{\mu\to\infty}\lim_{t\to\infty}\frac{1}{\mu t}\ln\mathbb{E}_{u}\Big[\exp\Big(\mu\sum_{i=1}^{t}u(i)\Big)\Big]=\lim_{t\to\infty}\frac{1}{t}\lim_{\mu\to\infty}\frac{1}{\mu}\ln\mathbb{E}_{u}\Big[\exp\Big(\mu\sum_{i=1}^{t}u(i)\Big)\Big]\label{eq:limit_change_2}\\
	&=\lim_{t\to\infty}\frac{1}{t}\lim_{\mu\to\infty}\frac{1}{\mu}\ln\Big(\Pr\{U(t)=t\cdot u_{\max}\}\exp\left(\mu t u_{\max}\right)\Big)\label{eq:limit_change_3}\\
	&=\lim_{t\to\infty}\frac{1}{t}\lim_{\mu\to\infty}\frac{1}{\mu}\ln\Big(\exp\left(\mu t u_{\max}\right)\Big)=\lim_{t\to\infty}\frac{1}{t}\left(tu_{\max}\right)=u_{\max}.\label{part_1_u_max}
\end{align}
\endgroup
In (\ref{eq:limit_change_2}), the first limit goes to $t\cdot u_{\max}$ for given $t$, and is primarily affected by the distribution of the energy arrival process \cite{kelly1996notes}.

\subsection{Proof of Proposition \ref{pro:mu_zero_and_infinity_2}}\label{app:proposition_mu_zero_and_infinity_2}
The proof of the proposition is similar to the proof in Appendix \ref{app:proposition_mu_zero_and_infinity}.

\subsection{Proof of Theorem \ref{theo:effective_capacity}}\label{app:theorem_effective_capacity}
In \cite[Chap. 7, Example 7.2.7]{chang_book}, it is shown for Markov modulated processes that $\frac{\Lambda(\theta)}{\theta}=\frac{1}{\theta}\ln\ssp\{M\Phi(\theta)\}=\frac{1}{\theta}\ln\ssp\{\Upsilon(\theta)\}$ where $\ssp\{\Upsilon(\theta)\}$ is the spectral radius of the matrix $\Upsilon(\theta)=M\times\Phi(\theta)$; $M$ is the transition matrix of the underlying Markov process and $\Phi(\theta)=\diag\{\phi_{0}(\theta),\cdots,\phi_{m-1}(\theta)\}$ is a diagonal matrix; components of which are the moment generating functions of the processes in $m$ states. We can see the rates (number of bits leaving the queue, $s(i)$) supported by the above channel model with the state transition model described in Section \ref{sec:energy_outage} as a Markov modulated process, and hence we can apply the setup considered in \cite{chang_book} immediately in our setting. Given that the battery moves to state 0 from any state, we have $s(i)=r(i)$ bits served from the data buffer in state 0 if $r(i)\leq \mathcal{I}(i)$ in the $i^{\text{th}}$ time frame. Otherwise, the effective transmission rate is zero, i.e., $s(i)=0$. Then, we have the following moment generating function in state 0:
\begingroup
\allowdisplaybreaks
\begin{align}
\phi_{0}(\theta)=&\mathbb{E}\left[\exp\left(\theta s(i)\right)|w(i)=0\right]=\mathbb{E}\left[\exp\left(\theta r(i)\textbf{1}\left[r(i)\leq\mathcal{I}(i)\right]\right)|w(i)=0\right]\\
=&\mathbb{E}\Big[\exp\left(\theta r(i)\right)\Pr\left\{r(i)\leq\mathcal{I}(i)\right\}+\Pr\left\{r(i)>\mathcal{I}(i)\right\}|w(i)=0\Big]\\
=&\frac{1}{\pi_{0}}\mathbb{E}_{p(i)}\Bigg[\int_{0}^{p(i)}\Big[\exp\left(\theta g(\xi(i))\right)\Pr\left\{g(\xi(i))\leq\mathcal{I}(i)\right\}+\Pr\left\{g(\xi(i))>\mathcal{I}(i)\right\}\Big]f_{\xi(i)}(\xi(i))d\xi(i)\Bigg]\nonumber\\
=&\frac{1}{\pi_{0}}\sum_{m=0}^{\infty}\pi_{m}\mathbb{E}_{p(i)}\Bigg[\int_{0}^{p(i)}\Big[\exp\left(\theta g(\xi(i))\right)\Pr\left\{g(\xi(i))\leq\mathcal{I}(i)\right\}+\Pr\left\{g(\xi(i))>\mathcal{I}(i)\right\}\Big]\nonumber\\
&\times f_{\xi(i)|w(i-1)=m}(\xi(i)|w(i-1)=m)d\xi(i)\Bigg]\nonumber\\
=&\frac{1}{\pi_{0}}\sum_{m=0}^{\infty}\pi_{m}\mathbb{E}_{p(i)}\Bigg[\int_{0}^{p(i)}\Big[\exp\left(\theta g(\xi(i))\right)\Pr\left\{\kappa(\xi(i))\leq|h(i)|^{2}\right\}+\Pr\left\{\kappa(\xi(i))>|h(i)|^{2}\right\}\Big]\nonumber\\
&\times f_{\xi(i)|w(i-1)=m}(\xi(i)|w(i-1)=m)d\xi(i)\Bigg].\label{eq:mgf_0}
\end{align}
\endgroup
Because the battery moves into state 0 when the amount of energy in the battery is less than what the transmitter demands, i.e., $e(i-1)+u(i)<p(i)$, the integral is taken from 0 to $p(i)$. The expression in (\ref{eq:mgf_0}) converges to a finite value with the increasing number of summands. Regarding the convergence, we refer to the methodology provided in Section \ref{subsubsec:constant_energy_rate}. Hence, it is enough to obtain a sum up to a certain value of $m$. Similarly, given that the battery is in state $j-1$ in the $(i-1)^{\text{th}}$ time frame and that it enters state $j$ in the $i^{\text{th}}$ time frame, we have $r(i)$ bits served in state $j$ if $r(i)\leq\mathcal{I}(i)$ and zero otherwise. Hence, we have the moment generating function in state $j$ as follows:
\begingroup
\allowdisplaybreaks
\begin{align}
\phi_{j}(\theta)&=\mathbb{E}\left[\exp\left(\theta s(i)\right)|w(i)=j\right]=\mathbb{E}\left[\exp\left(\theta r(i)\textbf{1}\left[r(i)\leq\mathcal{I}(i)\right]\right\}|w(i)=j\right]\\
=&\mathbb{E}\Big[\exp\left(\theta r(i)\right)\Pr\left\{r(i)\leq\mathcal{I}(i)\right\}+\Pr\left\{r(i)>\mathcal{I}(i)\right\}|w(i)=j\Big]\\
=&\frac{1}{\pi_{j}}\mathbb{E}_{p(i)}\Bigg[\int_{p(i)}^{\infty}\Big[\exp\left(\theta g(p(i))\right)\Pr\left\{\kappa(p(i))\leq|h(i)|^2\right\}+\Pr\left\{\kappa(p(i))>|h(i)|^2\right\}\Big]f_{\xi(i)}(\xi(i))d\xi(i)\Bigg]\nonumber\\
=&\frac{\pi_{j-1}}{\pi_{j}}\mathbb{E}_{p(i)}\Bigg[\int_{p(i)}^{\infty}\Big[\exp\left(\theta g(p(i))\right)\Pr\left\{\kappa(p(i))\leq|h(i)|^2\right\}+\Pr\left\{\kappa(p(i))>|h(i)|^2\right\}\Big]\nonumber\\
&\times f_{\xi(i)|w(i-1)=j-1}(\xi(i)|w(i-1)=j-1)d\xi(i)\Bigg]\\
=&\frac{1}{q_{j}}\mathbb{E}_{p(i)}\Bigg[\int_{p(i)}^{\infty}\Big[\exp\left(\theta g(p(i))\right)\Pr\left\{\kappa(p(i))\leq|h(i)|^2\right\}+\Pr\left\{\kappa(p(i))>|h(i)|^2\right\}\Big]\nonumber\\
&\times f_{\xi(i)|w(i-1)=j-1}(\xi(i)|w(i-1)=j-1)d\xi(i)\Bigg],
\end{align}
\endgroup
where the battery moves into state $j$ when the amount of energy in the battery is greater than or equal to what the transmitter demands, i.e., $e(i-1)+u(i)\geq p(i)$, the integral is taken from $p(i)$ to infinity. In addition, we observe that $\Upsilon(\theta)$ is the Leslie matrix \cite{Hansen}. Hence, the characteristic function of $\Upsilon(-\theta)$ is given in (\ref{characteristic_function}).

In order to analyze the roots of $z(\chi)$, we invoke the following theorem:
\begin{theo}[Cauchy's Theorem]
Let $z(\chi)=\chi^{n}-b_1\chi^{n-1}-\cdots-b_n$, where all the numbers $b_i$ are non-negative and at least one of them is non-zero. The polynomial $z(\chi)$ has a unique positive root $\chi^{*}$, and the absolute values of the other roots do not exceed $\chi^{*}$ \cite{Prasolov}.
\end{theo}

Note that $z(\chi)$ in (\ref{characteristic_function}) has coefficients that are non-negative, and at least one of them is non-zero. Therefore, there is one unique real positive root of $z(\chi)$, denoted by $\chi^{\star}$, which gives us the spectral radius of $\Upsilon(-\theta)$.

We know from \cite[Corollary 8.1.20]{Hansen} that any principal sub-matrix of $\Upsilon(-\theta)$, which is denoted by $\widehat{\Upsilon}(-\theta)$, has a spectral radius less than or equal to the spectral radius of $\Upsilon(-\theta)$, i.e., $\ssp\{\widehat{\Upsilon}(-\theta)\}\leq\ssp\{\Upsilon(-\theta)\}$ because $\Upsilon(-\theta)$ is non-negative matrix. Particularly, truncating the matrix, $\Upsilon(-\theta)$, to a finite size matrix $\Upsilon_{\alpha}(-\theta)$, i.e., from row number 1 to row number $\alpha$ and from column number 1 to column number $\alpha$, we will obtain an upper bound to the effective capacity, because $\ssp\{\Upsilon_{\alpha}(-\theta)\}\leq\ssp\{\Upsilon(-\theta)\}$ and $-\frac{1}{\theta}\ln\ssp\{\ssp\{\Upsilon_{\alpha}(-\theta)\}\}\geq-\frac{1}{\theta}\ln\ssp\{\Upsilon(-\theta)\}$. Hence, we reach the expression in (\ref{characteristic_function_upper_bound}), which is the characteristic function of $\Upsilon_{\alpha}(-\theta)$. Moreover, by increasing the truncated matrix size, the upper bound converges to the effective capacity. We can show this by noting that any defined truncated matrix is a principal sub-matrix of a bigger truncated matrix following the aforementioned definition of the truncated matrix.


\bibliographystyle{IEEEtran}
\bibliography{IEEEabrv,references}

\end{document}